\documentclass[5p]{elsarticle}

\usepackage{amssymb}

\journal{TIPP09 Proceedings in NIMA}

\newcommand{\C} {{\v{C}erenkov}}
\newcommand{\dream} {{\sc dream}}

\begin{document}

\begin{frontmatter}

\title{Dual-readout, Particle Identification, and 4th}

\author[Second]{Vito Di Benedetto}
\author[First]{John Hauptman\corref{cor1}}
\ead{hauptman@iastate.edu}
\cortext[cor1]{Corresponding author. Tel.: +1-515-451-0034} 
\author[Second]{Anna Mazzacane}

\address[First]     {Iowa State University, Ames, IA 50011, USA}
\address[Second]     { INFN and Dipartimento di Fisica, 
             via Lecce-Arnesano, 73100,   Lecce, Italy}

\begin{abstract}

The 4th detector is rich in particle identification measurements from the dual-readout
calorimeters, the cluster-timing tracking chamber, the muon spectrometer, and combinations of these systems.  In all, a total of 13 measurements contribute to the {\it identification of all partons of the standard model.}

\end{abstract}

%
%
%
%
%
%
\begin{keyword}

particle identification \sep dual readout calorimetry \sep 4th detector



\end{keyword}

\end{frontmatter}

The International Linear Collider (ILC) must be a near-perfect machine, and there must be near-perfect detectors whose measurement capabilities exceed those of the four excellent LEP detectors ({\sc aleph, delphi, l3, opal}) by large factors.  Critical to all physics is the identification of all standard model partons ($e, \mu, \tau, uds, c, b, t, W, Z, \gamma$, and $\nu$ by subtraction) from low energies to the highest energies.

The 4th detector \cite{4loi} is unique in several respects and makes particle identification measurements that are new in high energy physics.   In this brief paper,  we list the essential measurements and their resulting particle identifications in 4th, organized by the detectors that perform the essential measurements:

\medskip
\noindent {\bf dual-readout calorimeters}

\begin{figure}[h!]
\begin{center}
\includegraphics*[scale=0.45]{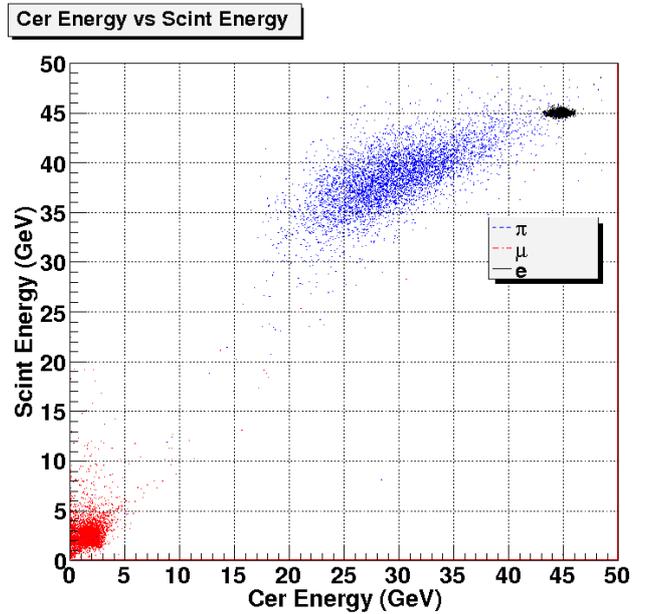}
\end{center}
\caption{\label{fig:pid2D}
The distributions of $S~ vs. ~C$ for $e^{\pm}$ (black points), $\pi^{\pm}$ (blue), and $\mu^{\pm}$ (red) particles at 45 GeV.  
}
\end{figure}

\medskip
\noindent {\it (i)  Measurement of scintillation (S) and \C (C) light} yields direct discrimination among $\pi^{\pm}, e^{\pm},$ and $\mu^{\pm}$ since these three species respond quite differently in $S$ and $C$.  The calorimeters are calibrated directly with electrons, and therefore an electron shower is characterized by $S \approx C$, whereas a pion (or any hadronic particle) is characterized by $S > C$ and, whenever momentum information ($p$) is available, also by $p > S > C$.  Muons in a fiber dual-readout calorimeter are characterized by $S \approx C + dE/dx \cdot \ell$, where $\ell$ is the muon pathlength in the calorimeter.  The basic dual-readout $S~ vs. ~C$ plot is shown in Fig. \ref{fig:pid2D}.

\begin{figure}[h!]
\begin{center}
\includegraphics*[scale=0.25]{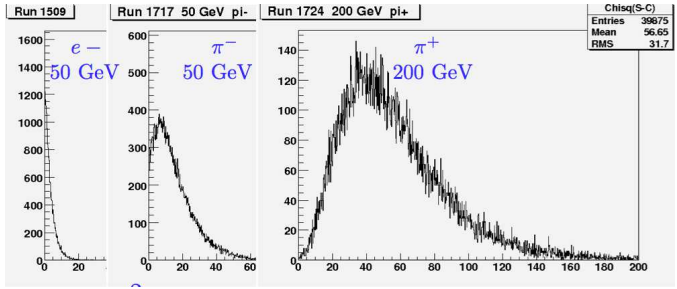}
\end{center}
\caption{\label{fig:chisqSC}
The distribution of $\chi^2_{S-C}$ for $e^-$ at 50 GeV, $\pi^-$ at 50 GeV, and $\pi^+$ at 200 GeV.
}
\end{figure}

\medskip
\noindent {\it (ii) Measurement of $S_k$ and $C_k$ channel-by-channel} is how the data are acquired and from which physics objects are pattern-recognized.   Fluctuations in $(S_k - C_k)$ channel-to-channel provide an independent measure of electromagnetic ({\sc em})  and hadronic identification.  For an {\sc em} shower, all of the channels $k$ that contribute to the shower individually have $S_k \approx C_k$, whereas a hadronic shower is characterized by very large spatial fluctuations in the local production and decay of $\pi^0 \rightarrow \gamma \gamma$.  A simple chi-squared statistic can be constructed to measure this discrimination:
\begin{displaymath}
 \chi^2_{S-C} = \sum_k [ (S_k - C_k)/\sigma_k]^2,
\end{displaymath}   
where $\sigma_k$ is the expected rms variation in $(S_k - C_k)$.  This $\chi^2$ is small for {\sc em} showers and large for hadronic showers.  This $\chi^2$ is shown for $e^-$ and $\pi^-$ at 50 GeV, and for $\pi^+$ at 200 GeV in Fig. \ref{fig:chisqSC}.

\medskip
\noindent {\it (iii) Measurement of $S$ and $C$ for isolated tracks} yields a unique identification of $\mu^{\pm}$ in the fiber dual-readout calorimeter.  It turns out that the \C ~angle is larger than the capture aperture of the \dream ~fibers and, therefore,  an approximately aligned muon passing through the fiber calorimeter has a zero \C ~ signal.  The muon generates scintillation light through ionization energy loss ($dE/dx$) as usual and, in addition, any bremsstrahlung or pair production from the muon within the massive absorber generates equal scintillation and \C ~light signals, $S \approx C$.  Therefore, in the quantity $S - C$ any amount of bremsstrah\-lung or pair production within the calorimeter cancels, and the difference is just the $dE/dx$ contribution to $S$.  In the \dream ~module, this is 1.1 GeV, and in a plot of  $(S - C) ~ vs. ~ (S+C)/2$ ~the $\mu^{\pm}$ are far removed from the $\pi^{\pm}$ tracks.

\medskip
\noindent {\it (iv) Measurement of the time-history of scintillation fibers} yields two identifications.  The first is a measure of the neutron content of a shower that is proportional to the scintillation signal at long times integrated over a larger volume, since the MeV neutrons liberated in nuclear break-up have velocities of about $v \sim 0.05c$.   The neutron signal is the scintillation light from the recoil protons in $np \rightarrow np$ elastic scatters, summed out to about $300$ ns.  The neutron fraction is $f_n \approx E_n/E_{\rm shower}$, which is strongly anti-correlated with the electromagnetic fraction, $f_{EM}$.

\begin{figure}[h!]
\begin{center}
\includegraphics*[scale=0.25]{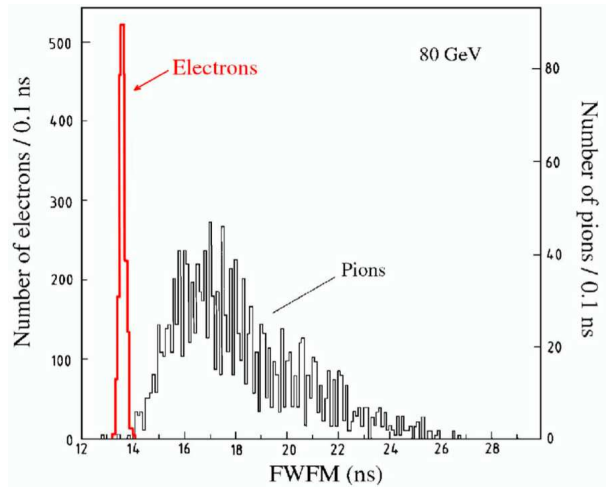}
\end{center}
\caption{\label{fig:spacal-e-pi}
Distriubution of the width of the scintillation pulse in time (full-width at 1/5-maximum) for $e^-$ and $\pi^-$ at 80 GeV.
}
\end{figure}

\medskip
\noindent {\it (v) Measurement of the time-duration of the scintillation signal} yields $e^{\pm}$ against $\pi^{\pm}$ discrimination since the fluctuations in the time-duration (measured as full-wdith at 1/5-maximum)  of {\sc em} showers is only 1-2 ns, whereas hadronic showers vary from 2-3 ns up to 20-25 ns.  This result was attained in the {\sc spacal} calorimeter,  Fig. \ref{fig:spacal-e-pi}.

\medskip
\noindent {\bf cluster-timing tracking chamber}

\begin{figure}[h!]
\begin{center}
\includegraphics*[scale=0.18]{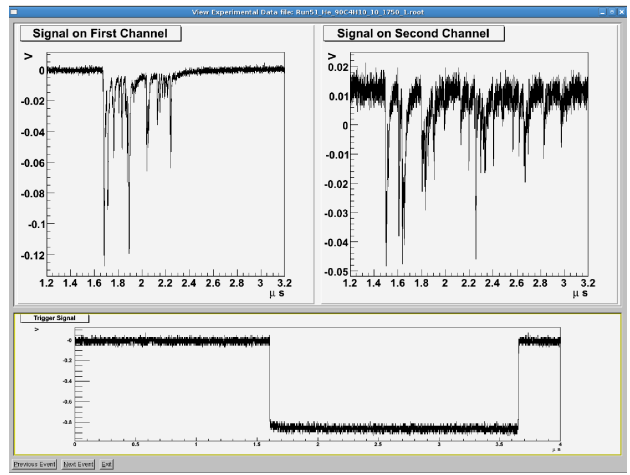}
\end{center}
\caption{\label{fig:clusters}
Clusters clocked at 2 GHz in the CluCou test for two different size tubes.  Bottom plot is the timing gate.
}
\end{figure}

\medskip \noindent {\it (vi) counting the number of ionization clusters} on a track is a Poisson measure of the specific ionization without Landau fluctuations.  The signal on each wire is read out with a GHz digitizer and individual clusters are timed and counted in the He-based gas that has  both a lower drift velocity and a lower number of clusters per unit length.  This counting has a low-side bias due to spatial cluster overlap and a high-side bias due to longitudinal diffusion in which a single cluster after a long drift may be counted as two clusters.  Neither of these effects, estimated as about 15\%,  are large enough to degrade this procedure, and we estimate a specific ionization measurement on tracks of length 1.2 meters at about 3\%.  Measured ionization clusters in CluCou are shown in Fig. \ref{fig:clusters}.

\medskip \noindent {\it (vii) $e-\gamma$ discrimination} depends on a well-measured trajectory in the tracker and a well-measured {\sc em} cluster in the calorimeter which are matched in both spatial coordinates and in energy-momentum. 

\medskip
\noindent {\bf dual-solenoid and muon spectrometer}

\medskip \noindent {\it (viii) $\mu^{\pm}$ tagging by energy balance} uses all detector systems of 4th:  the tracking system of pixel vertex chamber and cluster-timing chamber, the dual-readout calorimeters, and the muon spectrometer after the calorimeter and between the solenoids in a $B \sim 1.6$ T field with more than one meter of track length.   All tracks emerging from the back of the calorimeter are measured, and muons are tagged by energy balance between the central tracker momentum ($\sigma_p \sim {\rm few} \times 10^{-5} \cdot p^2$ GeV/c), the energy loss through radiation and ionization energy losses in the calori\-meters ($\sigma_E \sim 0.2 \sqrt{E}$ GeV), and the momentum measured in the muon spectrometer ($\sigma_p \sim 10^{-3} \cdot p^2$ GeV/c).  For a 50 GeV $\mu^{\pm}$ radiating 10 GeV in the calorimeter, this results in a pion rejection of about 30-50 against muons.  The combined $\mu^{\pm}$ tagging probabilities in the fiber calorimeter, {\it (i)} and {\it (iii)}, and corresponding $\pi^{\pm}$ rejection factors with this energy balance lead to extraordinary identification efficiency and purity for isolated $\mu^{\pm}$. 

\medskip
\noindent {\bf  {\sc bgo} crystal and fiber calorimeters} 

\medskip \noindent {\it (ix) Time-history of \C ~light} in both the crystal and fiber calorimeters provides a prompt signal that, if digitized at about 1 GHz, can be used to extract time-of-flight information on (primarily) {\sc em} showers with a time resolution of $\sigma_t \sim 0.3$ ns, or better.  It is not clear what physics is gained by this measurement, although exotic massive {\sc susy} objects have been speculated and, if they decay to light objects ($e, \mu, \tau, \gamma$), then these massive particles can be reconstructed by the arrive times and energies of their decay products.

\medskip \noindent {\it (x) Measurement of close-by {\sc em} and hadronic clusters} finds $\tau \rightarrow \rho \nu$ decays in which the $\pi^0 \rightarrow \gamma \gamma$ from the $\rho$ results in two close-by {\sc em} clusters in the finely segmented {\sc bgo} crystal calorimeter.  The charged $\pi^{\pm}$ from the $\rho$ can interact in the {\sc bgo} or pass through as a $mip$, but in either case its hadronic shower is absorbed and measured in the fiber calorimeter.   The two $\gamma$s are easily identified as {\sc em} showers since both $S \approx C$ and $\chi^2_{S-C} \approx 0$.  The $\pi^{\pm}$ is spatially separated from the two $\gamma$s in one-half the events (one-half $mip$ and one-half non-$mip$) and is further identified as hadronic by three identifications discussed above, {\it (i), (ii),} and {\it (iv)}, with almost no ambiguously since it is an isolated track from a one-prong $\tau$ decay. 

\begin{figure}[h!]
\begin{center}
\includegraphics*[scale=0.25]{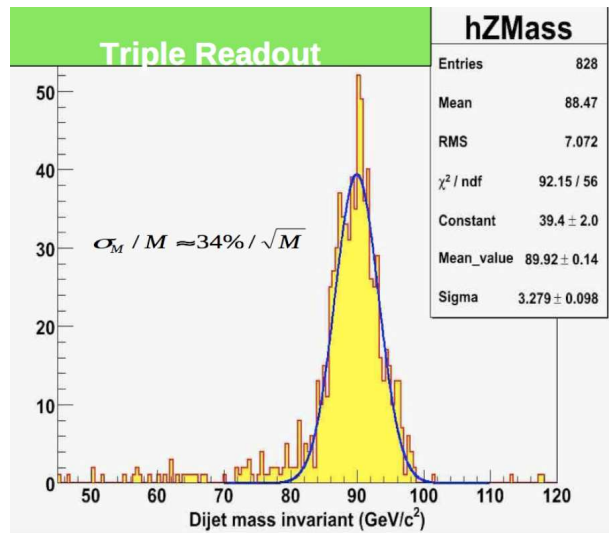}
\end{center}
\caption{\label{fig:M-di-jet}
The invariant mass distribution of di-jets.
}
\end{figure}

\medskip \noindent {\it (xi) Reconstruction and measurement of jets} is fundamental to ILC physics and leads directly to reconstruction of the hadronic decays $W \rightarrow jj$ and $Z^0 \rightarrow jj$, for which the resolution on the invariant mass of two jets achieved in 4th is characterized by $\sigma_M/M \approx 34\%/\sqrt{M}$.  This di-jet invariant mass resolution will be useful in the searches for massive exotic states decaying to $W$ and $Z$, and also directly to $q\bar{q}$.  This resolution is shown in Fig. \ref{fig:M-di-jet}.

\medskip \noindent {\it (xii) Measurement of high-precision {\sc em} energy} in the {\sc bgo} dual-readout calorimeter leads directly to the reconstruction of states at both high and low mass that  decay to photons:  $\pi^0 \rightarrow \gamma \gamma$ and $H^0 \rightarrow \gamma \gamma$.  This {\sc em} calorimeter has an energy resolution of $\sigma_E/E \approx 3\%/\sqrt{E}$.

\begin{figure}[hbt]
\begin{center}
\includegraphics*[scale=0.45]{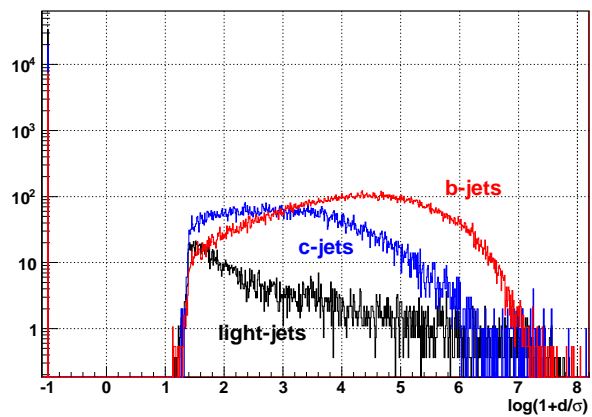}
\end{center}
\caption{\label{fig:btag}
Impact parameter distributions of secondary vertices of $uds, c$, and $b$ quarks from the {\sc rave} code.
}
\end{figure}

\medskip \noindent {\it (xiii) Measurement of track impact parameters} yields $uds$, $c$, and  $b$ quark secondary vertex discrimination as shown in Fig. \ref{fig:btag} from the {\sc rave} code.  

The 4th detector is a newly developed concept detector with several unique capabilities.  The calorimeters, both {\sc bgo} and fiber, have been thoroughly tested at CERN in extensive beam tests and published in 16 papers;  the cluster timing tracking chamber is based on the successful {\sc kloe} chamber augmented with GHz cluster timing electronics; and, the dual solenoid magnetic field configuration is calculated and its many benefits besides muon identification are described in Ref. \cite{4loi}.

\section*{Acknowledgments}

These results are supported by the U.S. Department of Energy under grant DE-FG02-01ER41155, and only possible due to the {\sc dream} collaboration and Fermilab and their support also from the U.S. Department of Energy.
%
%
%
%
%
%

\end{document}